\documentclass[preprint, onecolumn,preprintnumbers,tightenlines,floatfix, epsfig, showkeys]{revtex4}
\usepackage{dcolumn}
\usepackage{bm}
\usepackage{amsfonts}
\usepackage{amssymb}
\usepackage{amsmath}
\usepackage[ansinew]{inputenc}
\usepackage{graphicx}
\usepackage{color}
\usepackage[colorlinks]{hyperref}
\usepackage{eurosym}

\catcode`\=13
\def{$\bowtie$}
\DeclareInputText{128}{\euro} \DeclareInputText{165}{\yen}
\newtheorem{theorem}{Theorem}

\newtheorem{algorithm}[theorem]{Algorithm}

\newtheorem{definition}[theorem]{Definition}

\newtheorem{notation}[theorem]{Notation}

\input{tcilatex}

\begin{document}

\title{A Hybrid Quantum Encoding Algorithm of Vector Quantization for Image
Compression}
\author{Chao-Yang Pang$^{1,2}$}
\email{cyp_900@hotmail.com}
\author{Zheng-Wei Zhou$^{1}$}
\email{zwzhou@ustc.edu.cn}
\author{Guang-Can Guo$^{1}$}
\email{gcguo@ustc.edu.cn}
\affiliation{Key Laboratory of Quantum Information, University of Science and Technology
of China, Chinese Academy of Sciences, Hefei, Anhui 230026, China$^{1}$\\
College of Mathematics and Software Science, Sichuan Normal University,
Chengdu, Sichuan 610066, People's Republic of China$^{2}$}

\begin{abstract}
Many classical encoding algorithms of Vector Quantization (VQ) of image
compression that can obtain global optimal solution have computational
complexity $O(N)$. A pure quantum VQ encoding algorithm with probability of
success near 100\% has been proposed , that performs operations $45\sqrt{N}$
times approximately. In this paper, a hybrid quantum VQ encoding algorithm
between classical method and quantum algorithm is presented. The number of
its operations is less than $\sqrt{N}$ for most images, and it is more
efficient than the pure quantum algorithm.
\end{abstract}

\maketitle

\section{Introduction}

The two open research problem of vector quantization (VQ) are to design
real-time encoding algorithm and to design the codebook generation algorithm
that can generate approximately local optimal solution (finding global
optimal solution is NPC problem) \cite{1}. The property of local clustering
is revealed by Pang \cite{1} and it is used to design fast codebook
generation algorithm, that is faster than the famous LBG algorithm by a
factor of $4\symbol{126}13$ \cite{2}. But the other open problem, i.e., fast
encoding of VQ with global optimal solution has not been solved. The
encoding is described as \cite{1}:

Let codebook

\begin{equation*}
C=\{c[i]\mid \,0\leq i<N,c[i]\in R^{k},k\geq 1\}
\end{equation*}
\ \ 

For an arbitrary input vector $x\in R^{k}$ , we have to find the index $%
i_{0} $ such that distance $d(x,c[i_{0}])$ is minimum, where $R^{k}$ is $k-$
dimensional Euclidean Space. Vector $c[i]$ is called codevector.

Grover's algorithm \cite{3} is famous quantum search algorithm, and its
accuracy of rotation is improved recently \cite{4}. G.-L. Long proposes a
modified Grover's algorithm$\ $\cite{5}. Long's algorithm has the
probability of success 100\% even for the case that the size of database $N$
is very small. This property of Long's algorithm is suitable for quantum
image compression because $N$ is not a giant number in image compression in
general. Some hybrid search algorithms have also been proposed to improve
efficiency of algorithms \cite{6,7}, in which more physical resources are
used to commute the efficiency.\newline

Boyer, Brassard, H$\phi $yer and Tap propose an iteration algorithm named
BBHT algorithm in this paper, in which the Grover iteration is applied
repeatedly \cite{8}. BBHT algorithm (Boyer, Brassard, H$\phi $yer, Tap) \cite%
{8} is designed specially to solve the searching case that the number of
solutions is unknown.\newline

Quantum image compression is possible \cite{9,10,11}. Pang et.al., present a
Quantum Discrete Cosine Transformation (QDCT) for image compression \cite{9}%
. The method of QDCT can be applied to other transformation such as Fourier
transformation and classical output will be brought. Latorre presents a nice
and maybe hopeful quantum image compression method. Latorre's algorithm
requires that quantization method (scalar quantization or VQ) is
incorporated with it. Otherwise, it is not competitive with classical
methods \cite{10}. Pang, et.al., present a pure quantum VQ encoding
algorithm with probability of success near100\%, that performs operations $45%
\sqrt{N}$ times approximately \cite{11}.\newline
This paper presents a hybrid quantum VQ algorithm between classical method
and quantum algorithm to solve the \textit{open problem} of fast encoding of
VQ. And the number of its operations is less than $\sqrt{N}$ and its
probability of success is 100\% approximately.

\begin{notation}
$x$: \textit{input vector}
\end{notation}

\begin{notation}
$\delta _{0}=\min \{d(c[i],c[j])\mid \,i\neq j,0\leq i,j<N\}$
\end{notation}

\begin{notation}
$I$\textit{: whole encoding space}
\end{notation}

\begin{notation}
$S=\{x\,\mid \,d(x,c[i_{0}])<\frac{\delta _{0}}{2}\}$ and $T=\{x\,\mid
\,d(x,c[i_{0}])<\hat{\delta}\}$\textit{, where }$\hat{\delta}\geq \frac{%
\delta _{0}}{2}$.
\end{notation}

\begin{notation}
$\Omega (c[i])=\{c[j]\mid \,d(c[i],c[j])<2\hat{\delta},c[j]\in C\}$
\end{notation}

\begin{notation}
$Inf_{\Omega }=\min \{\mid \,\Omega \lbrack c[i]]\mid \mid \,0\leq i<N\}$
\end{notation}

All of the notations are diagrammatized in figure 2.\newline

Parameter $\delta _{0}$ can be calculated before encoding. The classical
data structure of $\Omega (c[i])$ can be constructed using classical method
before encoding (see figure 3).

\section{Quantum VQ Inequality Iteration and Quantum VQ Encoding}

Pang et.al., present a quantum representation of image to which database
technique is applied \cite{9}, and this quantum method is also used to
represent data of image in this paper.

\begin{definition}
Oracle $O_{d}$ :
\end{definition}

\begin{equation*}
\left\vert \delta \right\rangle \left\vert x\right\rangle \left\vert
i\right\rangle \left\vert c[i]\right\rangle \left\vert 0\right\rangle 
\overset{O_{d}}{\rightarrow }\left\vert \delta \right\rangle \left\vert
x\right\rangle \left\vert i\right\rangle \left\vert c[i]\right\rangle
\left\vert d(x,c[i])\right\rangle
\end{equation*}

$O_{d}$ is used to compute the distance $d(x,c[i])$ and $(O_{d})^{-1}$ is
the inverse transformation of it.

\begin{definition}
Oracle $O_{f}$ :
\end{definition}

\begin{equation*}
\left\vert \delta \right\rangle \left\vert x\right\rangle \left\vert
i\right\rangle \left\vert c[i]\right\rangle \left\vert
d(x,c[i])\right\rangle \overset{O_{f}}{\rightarrow }(-1)^{f(i)}\left\vert
\delta \right\rangle \left\vert x\right\rangle \left\vert i\right\rangle
\left\vert c[i]\right\rangle \left\vert d(x,c[i])\right\rangle
\end{equation*}%
, where function $f(i)$ is defined as

\begin{equation*}
f(i)=\left\{ 
\begin{tabular}{c}
\begin{tabular}{ccc}
$1$ & \ if & $d(x,c[i])<\delta $%
\end{tabular}
\\ 
\begin{tabular}{cc}
$0$ \ \ \ \ \ \ \ \  & otherwise%
\end{tabular}%
\end{tabular}%
\right.
\end{equation*}

\begin{definition}
quantum VQ inequality iteration $G_{inequality}$ :
\end{definition}

\begin{equation*}
G_{inequality}=(2\left\vert \xi \right\rangle \left\langle \xi \right\vert
-I)(U_{L})^{-1}(O_{d})^{-1}O_{f}O_{d}U_{L}
\end{equation*}%
, where $\left\vert \xi \right\rangle =\frac{1}{\sqrt{N}}\sum%
\limits_{i=0}^{N-1}\left\vert i\right\rangle $ and the unitary operation $%
U_{L}$ and $(U_{L})^{-1}$ is $LOAD$ operation that loads data into registers
from memory \cite{9}.

\begin{definition}
operation $U_{L}$:
\end{definition}

\begin{equation*}
\left\vert \delta \right\rangle \left\vert x\right\rangle \left\vert
i\right\rangle \left\vert 0\right\rangle \left\vert 0\right\rangle \overset{%
U_{L}}{\rightarrow }\left\vert \delta \right\rangle \left\vert
x\right\rangle \left\vert i\right\rangle \left\vert c[i]\right\rangle
\left\vert 0\right\rangle
\end{equation*}

\subsection{Sub-procedure 1: the Quantum VQ for $x\in S$}

\begin{algorithm}
Sub-procedure 1
\end{algorithm}

\begin{itemize}
\item[Step 1] Let $\delta =\frac{\delta _{0}}{2}$.

\item[Step 2] Generate the initial state$\mid \,\Psi _{0}>=\frac{1}{\sqrt{N}}%
\sum\limits_{i=0}^{N-1}(\left\vert \delta \right\rangle \left\vert
x\right\rangle \left\vert i\right\rangle \left\vert 0\right\rangle
\left\vert 0\right\rangle )$.

\item[Step 3] Perform the iteration $G_{inequality}$ $t=\left[ \frac{\pi }{4}%
\sqrt{N}\right] $ times:
\end{itemize}

\begin{equation*}
\mid \Psi _{t}>=(G_{inequality})^{t}\,\mid \,\psi _{0}>
\end{equation*}

\begin{itemize}
\item[Step 4] Observe the third register: let $i_{0}$ be the outcome.
Compute $d_{0}=d(x,c[i_{0}])$ classically.
\end{itemize}

\ \ If $d_{0}<\delta $ , the result is $i_{0}$ , otherwise go to
Sub-procedure 2.

\bigskip

If there exists a vector $c[i_{0}]\in C$ such that $d(x,c[i_{0}])<\frac{%
\delta _{0}}{2}$, the vector $c[i_{0}]$ is the closest codevector of $x\in S$
and it is the \textit{unique} closest codevector \cite{1}. And the iteration 
$G_{inequality}$ acts only on $N-$ dimensional subspace \cite{9,11}.
Therefore, by Grover's algorithm, sub-procedure 1 performs the operation $%
G_{inequality}$ $\left[ \frac{\pi }{4}\sqrt{N}\right] $ times approximately.
That is, sub-procedure 1 has time complexity $\left[ \frac{\pi }{4}\sqrt{N}%
\right] $.

\subsection{Sub-procedure 2: the Quantum VQ for $x\in (T-S)$ or $x\in (I-T)$}

\begin{algorithm}
Sub-Procedure 2
\end{algorithm}

\begin{itemize}
\item[Step 1] Initialize $m=1$ and $\lambda =\frac{6}{5}$.

\item[Step 2] We choose an experiential value $\hat{\delta}$ such that the
set $T$ can cover the whole encoding space $I$ approximately.

\item[Step 3] Choose $j$ uniformly at random among the nonnegative integers
not bigger than $m$.

\item[Step 4] Generate the initial state $\mid \Psi _{0}>=\frac{1}{\sqrt{N}}%
\sum\limits_{i=0}^{N-1}(\left\vert \delta \right\rangle \left\vert
x\right\rangle \left\vert i\right\rangle \left\vert 0\right\rangle
\left\vert 0\right\rangle )$

\item[Step 5] Apply $j$ iterations of the iteration $G_{inequality}$
starting from the state $\mid \Psi _{0}>$:
\end{itemize}

\begin{equation*}
\mid \Psi _{j}>=(G_{inequality})^{j}\,\mid \,\psi _{0}>
\end{equation*}

\begin{itemize}
\item[Step 6] Observe the third register: let $h$ be the outcome.

\item[Step 7] Perform classical computing $y_{0}=d(x,c[h])$.
\end{itemize}

\ \ If $y_{0}<\hat{\delta}$, classically full search the corresponding
neighborhood $\Omega (c[h])$ to obtain the closest codevector $c[i_{0}]$ ;
The result is $i_{0}$ and \textbf{exit}.

\begin{itemize}
\item[Step 8] Otherwise, set $m$ to $\min \{\lambda m,\sqrt{N}\}$ and go to
step 3.

\item[Step 9] \textbf{Encoding for} $I-T$ \textbf{:} If $y_{0}\geq \hat{%
\delta}$ or $i_{0}$ does not exist, apply classical method to encode.
\end{itemize}

\bigskip

We have conclusion that the closest codevector $c[i_{0}]$ of vector $x\in
T-S $ is included in the neighborhood $\Omega (c[h])$ of step 7 (see figure
1). The reason is that $d(x,c[j])\geq d(c[j],c[h])-d(c[h],x)>\hat{\delta}$
for an arbitrary $c[j]\notin \Omega (c[h])$.

The hybrid idea of sub-procedure 2 is that applying the iteration $%
G_{inequality}$ repeatedly to find a neighborhood $\Omega (c[h])$ which
comprises very few elements, then full search the neighborhood classically
to find the solution.

In addition, the set $I-T$ comprises very few vectors statistically and
classical method is applied to encode vector $x\in I-T$.

Figure 1 illustrates the hybrid idea of sub-procedure 2.

\bigskip

\bigskip 
\begin{tabular}{c}
{\LARGE Insert Figure 1 Here}%
\end{tabular}

\bigskip

The iteration $G_{inequality}$ is embedded in the BBHT algorithm in
sub-procedure 2. The case of no solution is handled by BBHT algorithm \cite%
{8}. The iteration $G_{inequality}$ acts only on $N-$ dimensional subspace 
\cite{9,11}. Thus, the number of performing the iteration $G_{inequality}$
in sub-procedure 2 is not bigger than $\frac{9}{4}\sqrt{\frac{N}{Inf_{\Omega
}}}$ when $Inf_{\Omega }\ll N$.

The four statistical properties of image for VQ are listed below (see figure
2):

\bigskip

\begin{itemize}
\item[Property 1] Almost feature vectors $x$ concentrate on themselves
centroids generally. That is, almost vectors $x\in I$ are included in set $S$%
. The solution is \textit{unique} for $x\in S$.

\item[Property 2] The set $T-S$ comprises the points of near edges of image
in practice. The solution is included in a small neighborhood $\Omega (c[h])$
for input vector $x\in T-S$.

\item[Property 3] The set $I-T$ comprises special points such as very detail
points or points at edges maybe.

\item[Property 4] Let $\mid S\mid :\mid T-S\mid :\mid I-T\mid =a:b:c$.
Statistically, we have (see figure 2)
\end{itemize}

\ \ \ \ \ $80\%\leq a\leq 99\%$, $1\%\leq b\leq 19\%$, and $c<1\%$

\bigskip

Therefore, the time complexity of the whole quantum encoding algorithm is
less than $\sqrt{N}$ statistically.

The phenomenon should be noticed, that the above properties are \textit{not}
powerful for classical methods to solve the open problem of the fast
encoding of VQ, by contrast, it \textit{is} powerful for quantum methods.%
\newline

Figure 2 shows these statistical properties of image.

\bigskip

\bigskip 
\begin{tabular}{c}
{\LARGE Insert Figure 2 Here}%
\end{tabular}

\bigskip

The hybrid algorithm is not only an image compression method, but also it
can be applied to image recognition \cite{12}.

\section{The Cost Physical Resource for Accessorial Data Structure}

The main accessorial resource is cost to save the classical data structure
of all neighborhoods $\Omega (c[i])$. And the data structure can be
constructed using adjacency list technique \cite{13} before encoding.
Adjacency list technique is a basic technique, and it is applied on image
compression usually \cite{1}$\ $. The adjacency list to save all $\Omega
(c[i])$ can be constructed by the following method \cite{13}:\newline

First, Allocate a single list to save all neighbors of each codevector $c[i]$%
, where $c[i]\in C$(i.e., centroid in figure 2) and $i=0,1,...,(N-1)$.

Second, All of the addresses of all single lists are saved in an array.

The structure of the adjacency list is diagrammatized in figure 3.\newline

It is easy to estimate the space complexity of the adjacency list$^{[13]}$.
The total number of the classical bits required to save the adjacency list
is approximately $\sum\limits_{i=0}^{N-1}(\mid \Omega (c[i])\mid +1)(\log
_{2}N+4\times 8)$ (bits). In general, $\mid \Omega (c[i])\mid $ is very
small (see figure 2).

Therefore, the space complexity is $O(N\log _{2}N)$ bits approximately and
it is very good.

It's very easy to realize the adjacency list for modern electronic computer 
\cite{13}. Therefore, it's a good tradeoff that costing $O(N\log _{2}N)$
classical bits to permute the running time by decreasing 45 factor than the
pure quantum algorithm presented in Ref.\cite{11}.

\bigskip

\bigskip 
\begin{tabular}{c}
{\LARGE Insert Figure 3 Here}%
\end{tabular}

\section{Conclusion}

VQ has high compression ratio and simple structure. However, the performance
comes at the cost of increased computational complexity. Fast encoding of VQ
is an open problem and is one of two key techniques of VQ. Many classical
encoding algorithms of VQ that can obtain global optimal solution have
computational complexity $O(N)$. In this paper, a hybrid quantum VQ encoding
algorithm between classical method and quantum algorithm is presented. The
number of its operations is less than $\sqrt{N}$ for most images. It's a
good tradeoff that costing $O(N\log _{2}N)$ classical bits to permute the
running time by decreasing 45 factor than the pure quantum algorithm in
Ref.[11].

\begin{acknowledgments}
The authors would like to thank prof.Valery Gorbachev for useful discussions
and comments, who is at St.-Petersburg State Uni. of AeroSpace
Instrumentation. The authors would like to very thank one of reviewers by
who the topic of section 3 of this paper is suggested. The authors thank
Prof. Z.-F. Han for his encouragement to us. The authors thank Y.-J. Han for
the very significant discussing with him especially. The authors thank Y-F
Huang, P.-X Chen, Y.-S. Zhang, J.-M. Cai and other colleagues for the
discussing with them. The authors thank Prof. J. Zhang and Prof. Z.-L. Pu,
of Sichuan Normal Univ. for his important support and help.
\end{acknowledgments}

\bigskip

\bigskip

\bigskip

\bigskip

\bigskip

\bigskip

\bigskip

\bigskip

\bigskip

\bigskip

\bigskip

\bigskip

\bigskip

\bigskip

\bigskip

\bigskip

\bigskip

\bigskip

\bigskip

\bigskip

\bigskip

\bigskip

\bigskip

\bigskip

\bigskip

\bigskip

\bigskip

\bigskip

\bigskip

\bigskip

\bigskip

\bigskip

\bigskip

\bigskip

\bigskip

\bigskip

\bigskip

\section{Figures}

\begin{figure}[h]
\begin{center}
\includegraphics[height=226pt,width=372pt]{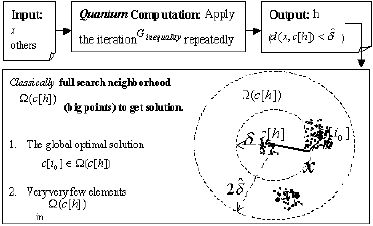}
\end{center}
\caption{The illustration of the hybrid idea of sub-procedure 2: \newline
{\protect\small 1. Embed the iteration }$G_{inequality}$ {\protect\small in
BBHT algorithm and apply the iteration repeatedly. This quantum method will
find a neighborhood }$\Omega (c[h])${\protect\small . \newline
2. The global optimal solution }$c[i_{0}]\in \Omega (c[h])${\protect\small ,
and there are very few elements in }$\Omega (c[h])${\protect\small .
Therefore, we can classically search }$\Omega (c[h])$ {\protect\small to
obtain solution.}}
\label{Fig1}
\end{figure}

\bigskip

\begin{figure}[tbp]
\begin{center}
\includegraphics[height=291pt,width=490pt]{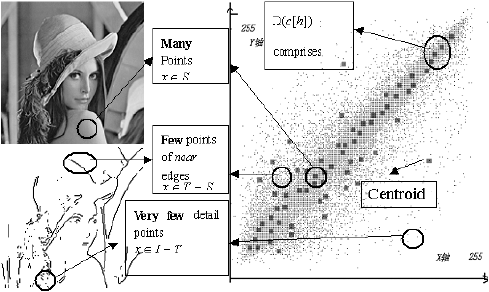}
\end{center}
\caption{The illustration of the properties of image that is applied to
design the hybrid algorithm:\newline
{\protect\small (Image Lena with size}$256\times 256${\protect\small \ is
divided into }$2\times 1${\protect\small \ image blocks to form vectors 
\protect\cite{1}.)}\newline
{\protect\small 1. Almost feature vectors }$x${\protect\small \ concentrate
on themselves centroids generally. And the solution is unique for }$x\in S$%
\newline
{\protect\small 2. The set }$T-S${\protect\small \ comprises the points of
near edges of image and it includes few points in practice. And the solution
is included in a small set }$\Omega (c[h])${\protect\small for }$x\in T-S$%
{\protect\small .} {\protect\small \newline
3. The set }$I-T${\protect\small \ comprises special points such as very
detail points or points at edges maybe. \newline
4. Statistically, }$\mid S\mid >>\mid T-S\mid >>>\mid I-T\mid $%
{\protect\small \newline
The above four statistical properties of image for VQ are applied to
accelerate the quantum algorithm: \newline
1. Sub-procedure 1 acts on the set }$S${\protect\small \ with time
complexity }$\left[ \frac{\protect\pi }{4}\protect\sqrt{N}\right] $%
{\protect\small \ approximately. \newline
2. Sub-procedure 2 acts on the set }$T-S${\protect\small \ with time
complexity less than }$\protect\sqrt{N}${\protect\small \ approximately. 
\newline
3. Classical full search algorithm acts on }$I-T${\protect\small . \newline
The phenomenon should be noticed, that these properties are not powerful for
classical methods to solve the open problem of the fast encoding of VQ, by
contrast, it is powerful for quantum methods.} }
\label{Fig2}
\end{figure}

\bigskip

\bigskip

\begin{figure}[tbp]
\begin{center}
\includegraphics[height=175pt,width=484pt]{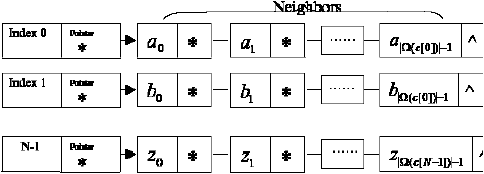}
\end{center}
\caption{Fig. 3 The illustration of the adjacency list to save all
neighborhoods $\Omega (c[i])$: \newline
Each single list saves all neighbors of $c[i]$. The space complexity of the
adjacency list is $O(N\log _{2}N)$ and it is very good. It's very easy to
realize the adjacency list for modern electronic computer. Therefore, it's a
good tradeoff that costing $O(N\log _{2}N)$ classical bits to permute the
running time by decreasing 45 factor than the pure quantum algorithm in Ref.%
\protect\cite{11}.}
\label{Fig3}
\end{figure}

\bigskip

\end{document}